\documentclass[showpacs,floatfix,superscriptaddress,showpacs,twocolumn,amssymb,amsfonts,prb,aps]{revtex4}% Physical Review Letters
\usepackage{longtable,graphicx,epsfig,dcolumn}
\begin{document}
\title{Wetting behavior \\
at the free surface of a liquid gallium-bismuth alloy:\\
An X-ray reflectivity study close to the bulk monotectic point.}
\author{P. Huber}
\address{Department of Physics, Harvard University, Cambridge MA 02138}
\author{O.G. Shpyrko}
\address{Department of Physics, Harvard University, Cambridge MA 02138}
\author{P.S. Pershan}
\address{Department of Physics, Harvard University, Cambridge MA 02138}
\author{H. Tostmann}
\address{Department of Chemistry, University of Florida, Gainesville FL 32611}
\author{E. DiMasi}
\address{Department of Physics, Brookhaven National Laboratory, Upton NY 11973}
\author{B.M. Ocko}
\address{Department of Physics, Brookhaven National Laboratory, Upton NY 11973}
\author{M. Deutsch}
\address{Department of Physics, Bar-Ilan University, Ramat-Gan 52900, Israel}
\date{\today }

\begin{abstract}
We present x-ray reflectivity measurements from the free surface of a liquid
gallium-bismuth alloy (Ga-Bi) in the temperature range close to the bulk
monotectic temperature $T_{mono}=222%
%TCIMACRO{\UNICODE[m]{0xb0}}%
%BeginExpansion
{{}^\circ}%
%EndExpansion
C$. Our measurements indicate a continuous formation of a thick wetting film
at the free surface of the binary system driven by the first order
transition in the bulk at the monotectic point. We show that the behavior
observed is that of a complete wetting at a tetra point of
solid-liquid-liquid-vapor coexistance.
\end{abstract}
\maketitle
\bigskip

Keywords: wetting, surface, liquid metals, gallium-bismuth, x-ray
reflectivity

\section{Introduction}

Interfacial phenomena at the surface of critical systems, particularly at
the surfaces of critical binary liquid mixtures, have attracted interest
ever since the seminal paper by J.W. Cahn \cite{cahn1977} demonstrating the
existence of a wetting line, starting at the wetting temperature, $T_{W}$,
below the critical point of demixing, $T_{crit}$. Initially, experimental
efforts were mainly focused on systems dominated by long-range van-der-Waals
interactions like methanol-cyclohexane and other organic materials.\cite
{Dietrich1988}\cite{Law2001} Only recently have experiments probed the
wetting behavior in binary metallic systems, most prominently in
gallium-lead (Ga-Pb) \cite{Chatain1996}, gallium-thallium (Ga-Tl) \cite
{Shim2001}, and gallium-bismuth (Ga-Bi) \cite{Nattland1996}. Those systems
allow the study of the influence of interactions characteristic of metallic
systems on the wetting behavior. By the same token, they allow one to obtain
information on the dominant interactions in metallic systems by measurements
of the thermodynamics and structure of the wetting films.

This paper reports measurements of the wetting behavior of Ga-Bi close to
its monotectic temperature $T_{mono}=222%
%TCIMACRO{\UNICODE[m]{0xb0}}%
%BeginExpansion
{{}^\circ}%
%EndExpansion
C$. We first present the bulk phase diagram of Ga-Bi and relate it to the
structures at the free surface (liquid-vapor interface) known from optical
ellipsometry \cite{Nattland1996} and recent x-ray reflectivity measurements.
\cite{Lei1996}\cite{Tostmann2000} The experimental setup is then described,
along with some basic principles of the x-ray reflectivity experiment and
the data analysis. Finally, our results and the conclusions emerging
therefrom are discussed.

\section{bulk and surface structure}

The bulk phase diagram of Ga-Bi has been measured by Predel using
calorimetric methods \cite{Predel1960}, and is shown in Fig. 1.
Assuming an
overall concentration of 70\% Ga, the following behavior is found: Below $%
T_{mono}$ (regime I) and at temperatures higher than the melting point of Ga
$T_{m}(Ga)=29.5%
%TCIMACRO{\UNICODE[m]{0xb0}}%
%BeginExpansion
{{}^\circ}%
%EndExpansion
C$, solid Bi coexists with a Ga-rich liquid phase. In this regime, previous
x-ray measurements have shown that a Gibbs-adsorbed Bi monolayer resides at
the free surface.\cite{Lei1996}\cite{Tostmann2000} For $T_{mono}<T<T_{crit}$
(regime II), the bulk phase separates into two immiscible phases, a high
density Bi-rich phase and a low density Ga-rich phase. The heavier Bi-rich
phase is macroscopically separated from the lighter Ga-rich phase and sinks
to the bottom of the sample pan. In temperature regime II, the high density,
Bi-rich phase wets the free surface by intruding between the low density
phase and the Bi monolayer, in defiance of gravity.\cite{Tostmann2000}
Considering that pure Bi has a significantly lower surface tension than pure
Ga, the segregation of the Bi-rich phase at the surface is not too
surprising. In fact, the thickness of the wetting layer in such a geometry
is believed to be limited only by the extra gravitational potential energy
paid for having the heavier phase at the top \cite{DeGennes1985}. In regime
III, above the consolute point, the bulk is in the homogeneous phase and
only a Gibbs-adsorbed Bi monolayer has been found at the free surface.\cite
{Huber2001}
\begin{figure*}[tbp]
\epsfig{file=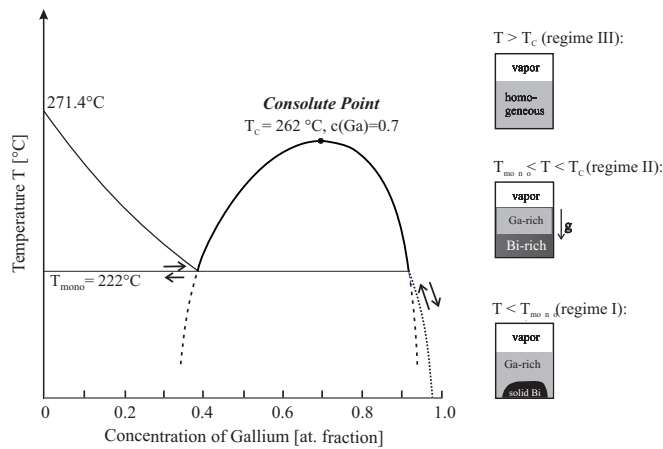, angle=0, width=1.0\columnwidth}
\caption{\label{fig:Fig1}Bulk Phase diagram of Ga-Bi (Ref.
\cite{Predel1960}). Bold solid line: liquid-liquid coexistence,
dashed line: liquid-liquid coexistence metastable, dotted line:
liquid-solid coexistence. On the right: Schematic bulk behavior in
the different temperature regimes.}
\end{figure*}

Despite extensive experimental efforts to relate the bulk phases
to the corresponding surface phases, two important question remain
unanswered: What kind of transitions take place at the free
surface between regime I and regime II and between regime II and
regime III. Here, we focus on the transition between regime I and
regime II. Note that these temperature regimes are separated by
$T_{mono}$, a temperature characterized by the coexistence of the
solid Bi, the Bi-rich liquid, the Ga-rich liquid and the vapor.
$T_{mono}$ is therefore a solid-liquid-liquid-vapor tetra point,
at which the solidification of pure bulk Bi takes place. The
question is, then, how is this first order bulk transition related
to the changes at the surface?

\section{Experiment}

The Ga-Bi alloy was prepared in an inert-gas box using metals with purities
greater than 99.9999\%. A solid Bi ingot was placed in a Mo pan and oxide
formed at the surface was scraped away. Liquid Ga was then added to
completely cover the Bi ingot. The sample with an overall Ga amount of 70
at\% was transferred through air into an ultrahigh vacuum chamber. After a
one day of bake-out, a pressure of 10$^{-10}$torr was achieved and the
residual oxide on the liquid sample was removed by sputtering with Ar$^{+}$
ions, at a sputter current of a 25 microamps and a sputter voltage of 2 kV.

To avoid temperature gradients between the bulk and the surface induced by
thermal radiation, a temperature-controlled radiation shield was installed
above the sample. The temperature was measured with thermocouples mounted in
the bottom of the sample pan and on the radiation shield. The sample pan and
the radiation shield temperature were controlled by Eurotherm temperature
controllers yielding a temperature stability of $\pm 0.05%
%TCIMACRO{\UNICODE[m]{0xb0}}%
%BeginExpansion
{{}^\circ}%
%EndExpansion
C$.

Surface-specific x-ray reflectivity measurements were carried out using the
liquid surface reflectometer at beamline X22B at National Synchrotron Light
Source with an x-ray wavelength $\lambda =1.54\AA .$ Background and bulk
scattering were subtracted from the specular signal by displacing the
detector out of the reflection plane. The intensity $R(q_{z})$, reflected
from the surface, is measured as a function of the normal component $q_{z}$
of the momentum transfer and yields information about the surface-normal
structure of the electron density as given by

\[
R(q_{z})=R_{F}(q_{z})\mid \Phi (q_{z})\mid ^{2}\exp [-\sigma
_{cw}^{2}q_{z}^{2}]\text{ \ \ (1),}
\]

\noindent where $R_{F}(q_{z})$ is the Fresnel reflectivity from a flat,
infinitely sharp surface, and $\Phi (q_{z})$ is the Fourier transform of the
local surface-normal density profile $<\widetilde{\rho }(z)>$\cite
{Pershan1984}:

$\bigskip $%
\[
\Phi (q_{z})=\frac{1}{\rho _{\infty }}\int dz\frac{d<\widetilde{\rho }(z)>}{%
dz}\exp (iq_{z}z)\text{ \ \ (2),}\
\]

\noindent with the bulk electron density $\rho _{\infty }$ and the critical
wave vector $q_{crit}$. The exponential factor in Eq. (1) accounts for
roughening of the intrinsic density profile $<\widetilde{\rho }(z)>$ by
capillary waves:

$\smallskip $%
\[
\sigma _{cw}^{2}=\frac{k_{B}T}{2\pi \gamma }\ln (\frac{q_{\max }}{q_{res}})%
\text{ \ \ (3),}
\]
\

\noindent where $\gamma $ is the macroscopic surface tension of the free
surface, and $\sigma _{cw}$ is the roughness due to thermally excited
capillary waves (CW) . The CW spectrum is cut off at small $q_{z}$ by the
detector resolution $q_{res}=0.03$\AA $^{-1}$ and at large $q_{z}$ by the
inverse atomic size $a$ , $q_{max\bigskip }\approx \pi /a$.\cite{Ocko1994}.

In Expression (1) the validity of the Born approximation is tacitly assumed.
Since the features in $R/R_{F}$ characteristic of the thick wetting film
appear close to the critical wavevector $q_{c}=0.049$\AA $^{-1}$ of the
Ga-rich subphase, where the Born approximation is no longer valid, we had to
resort to Parratt's dynamical formalism \cite{Parratt1954} for $q_{z}$ less
than $0.25\text{\AA }^{-1}$. Details of this analysis will be reported
elsewhere.\cite{Huber2001}

\section{Results}

Fig. 2a shows the reflectivity at three temperatures: well below $T_{mono}$
at $T=205%
%TCIMACRO{\UNICODE[m]{0xb0}}%
%BeginExpansion
{{}^\circ}%
%EndExpansion
C$ (regime I), well above $T_{mono}$ at $T=222.5%
%TCIMACRO{\UNICODE[m]{0xb0}}%
%BeginExpansion
{{}^\circ}%
%EndExpansion
C$ (regime II) and at an intermediate temperature $T=220%
%TCIMACRO{\UNICODE[m]{0xb0}}%
%BeginExpansion
{{}^\circ}%
%EndExpansion
C.$ Due to the loss of phase information (Eq. 1) the interesting
electron density profiles cannot be obtained directly from the
measured reflectivity. One has to resort to the widely accepted
procedure of adapting a physically motivated model for the
electron density profile and fitting its Fourier transform to the
experimentally determined $R/R_{F}$. The resulting density
profiles for the different temperatures are depicted in Fig. 2b
and the corresponding fits to $R/R_{F}$ can be found as solid
lines in Fig. 2a.

At $T=205%
%TCIMACRO{\UNICODE[m]{0xb0}}%
%BeginExpansion
{{}^\circ}%
%EndExpansion
C$ $R/R_{F}$ (Fig. 2a, diamonds) shows a pronounced maximum centered around $%
q_{z}=0.8\AA ^{-1}$. This maximum is indicative of a high electron density
at the surface of the alloy and a previous analysis shows that it is
compatible with the segregation of a monolayer of pure Bi at the
liquid-vapor interface of the bulk alloy.\cite{Tostmann2000}

The normalized reflectivity at $T=222.5%
%TCIMACRO{\UNICODE[m]{0xb0}}%
%BeginExpansion
{{}^\circ}%
%EndExpansion
C$ (Fig. 2a, solid circles) shows two peaks (Kiessig fringes) at
low $q_{z}$ characteristic of a Bi-rich wetting film $\approx
50\AA $\ thick. This thickness is in good agreement with
ellipsometric measurements on Ga-Bi alloy surfaces in the same
temperature regime.\cite{Nattland1996} The interfacial roughness
between the Bi-rich and the Bi-poor phase, indicated
by the decay of the Kiessig fringes, is $\approx 12\AA $. Additionally, $%
R/R_{F}$ continues to exhibit an increased intensity around $q_{z}=0.8\AA
^{-1}$: The monolayer of pure Bi is still present at the surface in regime
II, while the Bi-rich wetting film has intruded between it and the Ga-rich
subphase.

\begin{figure}[tbp]
\epsfig{file=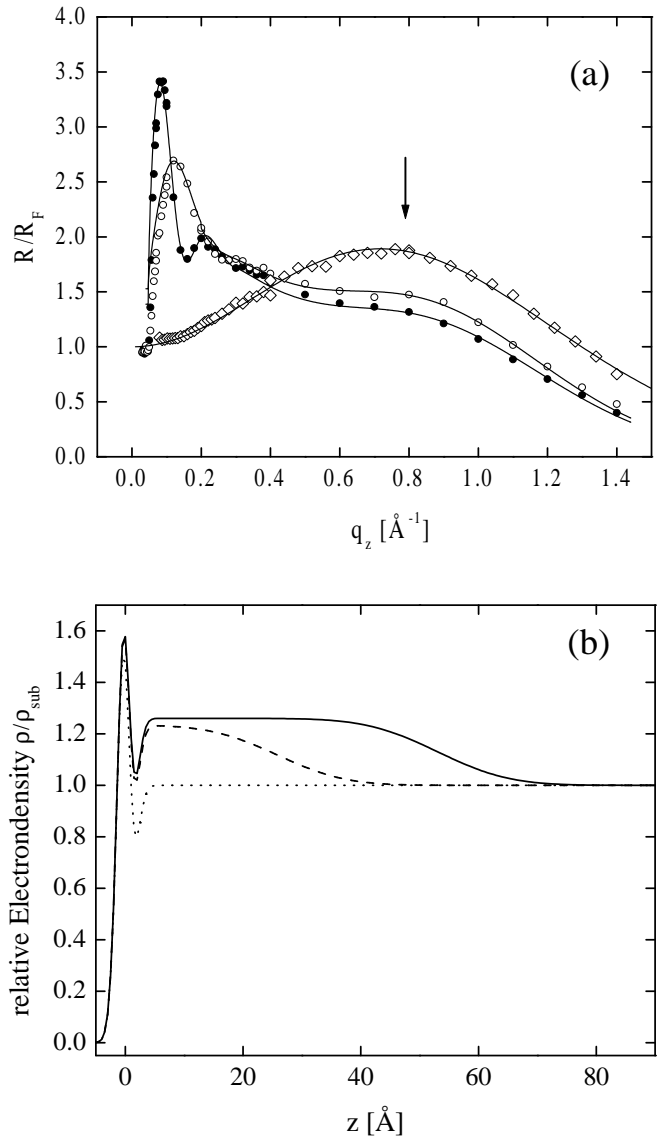, angle=0, width=1.0\columnwidth}
\caption{\label{fig:Fig2} (a) Fresnel normalized x-ray
reflectivity $R/R_{F}$ from the surface
of Ga-Bi: (diamonds) $T=205%
%TCIMACRO{\UNICODE[m]{0xb0}}%
%BeginExpansion
{{}^\circ}%
%EndExpansion
C$, (open circles) $T=220%
%TCIMACRO{\UNICODE[m]{0xb0}}%
%BeginExpansion
{{}^\circ}%
%EndExpansion
C$, (solid circles) $T=222.5%
%TCIMACRO{\UNICODE[m]{0xb0}}%
%BeginExpansion
{{}^\circ}%
%EndExpansion
C$. solid lines: fit to $R/R_{F}$ with density profiles depicted
in Fig. 2b. The arrow indicates the $q_{z}$-position for the
temperature-dependent reflectivity measurements at fixed
$q_{z}=0.8\AA ^{-1}$ (T-scan) shown in Fig. 3.\\
(b): Electron density profiles for $T=205%
%TCIMACRO{\UNICODE[m]{0xb0}}%
%BeginExpansion
{{}^\circ}%
%EndExpansion
C$ (dotted line), $T=220%
%TCIMACRO{\UNICODE[m]{0xb0}}%
%BeginExpansion
{{}^\circ}%
%EndExpansion
C$ (dashed line)\ and $T=222.5%
%TCIMACRO{\UNICODE[m]{0xb0}}%
%BeginExpansion
{{}^\circ}%
%EndExpansion
C$ (solid line), normalized to the bulk electron density of the
Ga-rich subphase.}
\end{figure}

The reflectivity at the intermediate temperature $T=220.0%
%TCIMACRO{\UNICODE[m]{0xb0}}%
%BeginExpansion
{{}^\circ}%
%EndExpansion
C$ (Fig. 2a, open circles) still exhibits a peak at low $q_{z}$.
However,
compared to the peak at $T=222.5%
%TCIMACRO{\UNICODE[m]{0xb0}}%
%BeginExpansion
{{}^\circ}%
%EndExpansion
C,$ it is now shifted to higher $q_{z}$. The corresponding
electron density profile, depicted in Fig. 2b as the dashed curve,
indicates a highly diffuse, thin film of a Bi enriched phase
between the monolayer and the Ga-rich subphase.

To obtain further information about the temperature dependent
transition between the different surface regimes, we performed
temperature-dependent x-ray reflectivity measurements at a fixed
$q_{z}$ in a temperature range close to T$_{mono}$. These
measurements are referred to in the following as T-scans. As can
be seen in Fig. 2a the reflectivities in the two regimes differ
most distinctly in the low $q_{z}$ part, and somewhat less, but
still significantly in the higher $q_{z}$-part. Due to the high
sensitivity of the reflectivity to temperature dependent sample
height changes at low $q_{z}$, we carried out measurements at a
relatively high $q_{z}=0.8\AA ^{-1}$.

A T-scan while cooling (heating) with a cooling (heating) rate of 1$%
%TCIMACRO{\UNICODE[m]{0xb0}}%
%BeginExpansion
{{}^\circ}%
%EndExpansion
C$/hour between $226%
%TCIMACRO{\UNICODE[m]{0xb0}}%
%BeginExpansion
{{}^\circ}%
%EndExpansion
C$ and $210%
%TCIMACRO{\UNICODE[m]{0xb0}}%
%BeginExpansion
{{}^\circ}%
%EndExpansion
C$ is shown in Fig. 3. For temperatures higher than $T_{mono}$ the
reflectivity stays constant. Upon $T$ reaching $T_{mono}$ the
intensity
starts rising continuously. When reaching $210%
%TCIMACRO{\UNICODE[m]{0xb0}}%
%BeginExpansion
{{}^\circ}%
%EndExpansion
C$ the x-ray reflectivity has increased by about 30\% as compared
to its value above $T_{mono}$ - as one would expect from Fig. 2a
for the change from regime I to regime II at that
$q_{z}$-position. While heating, we observed the inverse behavior,
the intensity decreases until it reaches its lowest value at
$T_{mono}$. No hysteresis was observed between cooling and
heating for cooling (heating)\ rates smaller than $2%
%TCIMACRO{\UNICODE[m]{0xb0}}%
%BeginExpansion
{{}^\circ}%
%EndExpansion
C$ per hour. We also carried out one temperature scan with a cooling rate of
10%
%TCIMACRO{\UNICODE{0xb0}}%
%BeginExpansion
${{}^\circ}$%
%EndExpansion
C per hour. This showed a hysteresis of $4%
%TCIMACRO{\UNICODE[m]{0xb0}}%
%BeginExpansion
{{}^\circ}%
%EndExpansion
C$ and also the formation of a solid film at the surface. We attribute this
to the expected diffusion-limited growth and the corresponding long
equilibration times in such wetting geometries\cite{Lipowsky1986} and to a
temperature gradient between bulk and surface rather than to an intrinsic
feature of the transition.

\begin{figure}[tbp]
\epsfig{file=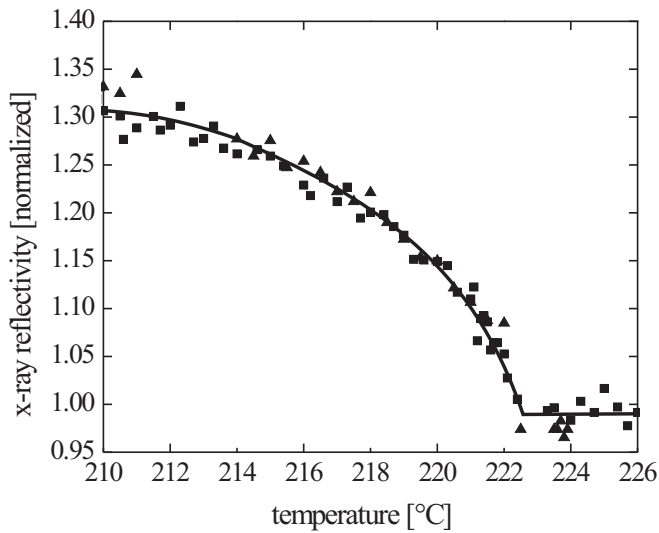, angle=0, width=1.0\columnwidth}
\caption{\label{fig:Fig3}temperature-dependent x-ray reflectivity
at $q_{z}=0.8\AA ^{-1}$
normalized to $R/R_{F}$ at $T=222.5%
%TCIMACRO{\UNICODE[m]{0xb0}}%
%BeginExpansion
{{}^\circ}%
%EndExpansion
C$. (squares) decreasing temperature, (triangles) increasing
temperature. line: guide for the eye.}
\end{figure}

\section{Conclusions}

The behavior of the reflectivity in the T-scan shows that the Bi-rich
wetting film forms while approaching $T_{mono}$ from below and vanishes in
the same way on cooling, i.e. without hysteresis. Hence, the surface
sensitive x-ray reflectivity measurements suggest that the first order bulk
transition at the monotectic point drives a continuous structural transition
at the surface. A similar behavior has been reported for Ga-Pb, a system
with an analogous phase diagram (monotectic point + consolute point ), by
Wynblatt and Chatain \cite{Chatain1996}. The topology of the phase diagram
of Ga-Bi forces the approach of the liquid-liquid coexistence line at $%
T_{mono}$ from off coexistence while heating. Conversely, it
enforces a path leading off coexistence while cooling below
$T_{mono}$- see arrows in Fig. 1. Therefore, as Dietrich and
Schick \cite{Dietrich1997} have pointed out, one measures along a
path that probes a complete wetting scenario (according to the
wetting nomenclature). Since the path ends in the monotectic point
with its four phase coexistence, the phenomenon is readily
described as a complete wetting at a tetra point of four-phase
coexistence. It should, however, be noted that $T_{mono}$ is not
the wetting temperature, $T_{w}$ of the liquid-liquid system.
$T_{w}$ is defined as the temperature where the formation of the
wetting film can be observed on coexistence between the two
liquids. This temperature has to be below $T_{mono}$, but it would
be experimentally accessible only if the pure Bi can be
supercooled sufficiently to reach the liquid-liquid coexistence,
albeit in a metastable condition.\newline

The observed transition at the surface is closely related to triple point
wetting phenomena \cite{Pandit1983} which have been extensively studied for
one component systems like Krypton on graphite \cite{Zimmerli1992}. There,
one observes that the liquid wets the interface between substrate and vapor,
while the solid does not. Hence, there is a wetting transition pinned at the
bulk triple point $T_{3}$. In the triple point wetting scenario one walks
off the liquid-vapor coexistence line following the solid-vapor sublimation
line.

Another aspect of the wetting behavior close to the monotectic point is the
possible nucleation of solid Bi at the surface while cooling below $T_{mono}$%
. Systematic studies on that aspect were reported only recently for
different Ga-Bi alloys, concluding that Ga-Bi shows a phenomenon that can be
described as surface freezing. The free surface acts as substrate for
wetting of the liquid by the forming solid Bi \cite{Wang2000}\cite
{turchanin2001}. We did also observe the formation of solid Bi at the
surface while cooling below $200%
%TCIMACRO{\UNICODE[m]{0xb0}}%
%BeginExpansion
{{}^\circ}%
%EndExpansion
C$, but we could not study this effect in detail, since the forming solid
film destroyed the flat surface of the liquid sample hampering further
reflectivity measurements.

A more detailed experimental study of the wetting transition at $T_{mono}$
has already been performed \cite{Huber2001}. It has revealed the evolution
of the thick film via intermediate film structures dominated by strong
concentration gradients - as reported here for $T=220%
%TCIMACRO{\UNICODE[m]{0xb0}}%
%BeginExpansion
{{}^\circ}%
%EndExpansion
C$. This behavior is in agreement with density functional calculations for
wetting transitions of binary systems at hard walls \cite{Davis1996}, which
also reveal concentration gradients. Conversely, the study shows that the
liquid vapor interface of a metal system acts not only as a hard wall for
one component systems \cite{Harris1987} forcing the ions into ordered layers
parallel to that surface \cite{Magnussen1995},\cite{Regan1995}, but also
affects other structure and thermodynamic phenomena at the surface, e.g. the
wetting behavior in a binary liquid metal system discussed here.

\section{Acknowledgments}

This work was supported by the U.S. DOE Grant No. DE-FG02-88-ER45379,
National Science Foundation Grant NSF-DMR-98-72817, and the U.S.-Israel
Binational Science Foundation, Jerusalem. Brookhaven National Laboratory is
supported by U.S. DOE Contract No. DE-AC02-98CH10886. P. Huber acknowledges
support from the Deutsche Forschungsgemeinschaft.

\end{document}